\documentclass[12pt,a4]{article}

\usepackage{amsmath, amssymb}
\usepackage{graphicx}
\usepackage{booktabs}

\usepackage{amsmath}

\usepackage[a4paper, total={16cm, 24cm}]{geometry}

\usepackage{hyperref}

\usepackage{fouriernc}        

\usepackage[auth-lg]{authblk} 

\usepackage[font={footnotesize,sf},labelfont=bf]{caption}

\usepackage[numbers,sort]{natbib}
\usepackage{tikz}
\usetikzlibrary{arrows.meta,positioning,fit,shapes.geometric,calc}

\title{A Unified Framework for Structured Flow Modeling:
  From Representation to Verification and Model Discovery}

\author{\Large Diego Casadei}
\affil{\normalsize\emph{Fly High Engineering AG}, Badenerstrasse 13, 5200 Brugg AG, Switzerland}

\date{June 12, 2026} 

\usepackage{xcolor}

\renewcommand{\vec}[1]{\ensuremath{\boldsymbol{#1}}}   
\newcommand{\mat}[1]{\ensuremath{\mathsf{#1}}}         
\newcommand{\de}{\ensuremath{\text{d}}}                

\begin{document}

\maketitle

\begin{abstract}
Many dynamical systems can be described in terms of structured flows combining source/sink behavior, cyclic dynamics, and topology-constrained transport. These features arise across a wide range of physical, engineered, and data-driven systems. The objective of this work is to establish a unified perspective on such systems, to identify modeling approaches that balance expressivity, interpretability, computational complexity, and data requirements, and to investigate how highly expressive models can be used to uncover the dominant mechanisms underlying observed dynamics.

Starting from the Helmholtz-Hodge decomposition of continuous vector fields, we review the recently proposed Graph Vector Field (GVF) framework and its discrete representation on simplicial complexes. We then introduce a hierarchy of alternative approaches, including parametric conditional models, linear graph dynamical systems, and reduced Hodge representations. Finally, we propose a verification and validation methodology based on benchmark datasets from well-understood physical systems and on systematic model-reduction and ablation studies.

The resulting family of structured-flow models within a common framework, ranging from low-dimensional parametric representations to full GVF formulations, supports a diagnostic methodology in which gradient, curl, harmonic, and topological contributions are systematically assessed through ablation studies. This process enables the identification of dominant mechanisms underlying the observed dynamics and guides the construction of simplified models tailored to the available data and operational constraints.

By separating structural verification, behavioral verification, and domain-specific validation, the proposed approach provides a foundation for scalable and interpretable analysis of complex dynamical systems across multiple application domains.
\end{abstract}

\newpage
\tableofcontents
\newpage

\section{Introduction}\label{sec-intro}

Many dynamical systems can be described in terms of flows that exhibit
source/sink behavior, cyclic dynamics, and transport constrained by
the underlying topology. Such structured flows arise across a wide
range of domains, including physical systems governed by conservation
laws, engineered networks with predefined connectivity, and
data-driven systems in which interactions must be inferred. Despite
their diversity, these systems share common mathematical features that
suggest the possibility of a unified modeling framework.

These systems may be grouped in different classes, accoding to how the flow is constrained and represented:
\begin{enumerate}
  \item Network-constrained systems with predefined topology;
  \item Continuous physical systems requiring discretization;
  \item Data-driven systems where topology is inferred;
  \item Engineered multi-agent systems with dynamically evolving topology.
\end{enumerate}

The first class includes network-constrained systems, which are
discrete by nature and have predefined connections, like pipelines
(water, gas, oil), electrical grids, traffic and transport networks,
communication networks and the whole Internet, sensor networks, and
possibly also supply chains.  While their topology may change, this is
a local perturbation affecting a minor fraction of the overall
network.  Moreover, flow laws may be known from physics, like Kirchoff
laws for electrical networks, or be inferred from broad principles,
acting like conservation laws (e.g.\ cars don't disappear between to
crossroads).

The second class includes continuous systems, which need to be
discretized for computational purposes to obtain e.g.\ a tractable
finite element model (FEM).  Examples include fluid dynamics (oceans,
atmosphere, climate), heat transfer (conduction, convection, and
irradiation), plasma physics (space weather, tokamak), elastic systems
(stress, strain), and transport in porous media.  The dynamics of such
systems is naturally described by partial differential equations
(PDEs) and their mapping onto a discrete mesh is not unique, as it is
well known by the practicioners, who tune FEM nodes to achieve the
most reliable results.

The third class of systems includes networks that arise from the data
themselves, rather than being the result of a deterministic
optimization, or emerging from intrinsic topological properties.
Examples include biological systems (metabolic networks, physiology),
brain connectivity, epidemiology, social systems, human interaction
networks and financial flows.  When topology emerges from the data,
the resulting graph needs to be interpreted, which is not always a
simple task.  For example, an edge connecting two nodes may represent
statistical or causal relationships.

The fourth class may include system described by a topology that is
actively changing and controlled, like robot or drone swarms, fleets
of autonomous vehicles, and distributed control systems.  In these
cases, the topology is part of the control problem and links appear
and disappear intentionally.  Feedback loops are the central focus of
such systems.
Table \ref{tab-classes} provides a compact comparison across the four classes.

\begin{table}[ht]
\caption{Classes of systems exhibiting structured flows.}
\label{tab-classes}
  \centering
  \small
\begin{tabular}{lll}
\toprule
Class                & Description                       & Examples \\
\midrule
Network-constrained  & Fixed or slowly varying topology  & Pipelines, power grids, traffic \\
Continuous systems   & PDE-based, require discretization & Fluids, climate, plasma \\
Data-driven networks & Topology inferred from data       & Biology, epidemiology, finance \\
Multi-agent systems  & Dynamic, controlled topology      & Swarms, robotics \\
\bottomrule
\end{tabular}
\end{table}

Whereas all these classes include gradient-driven, curl-driven, and
topology-constrained components, they do not exhibit all three
components equally.  For example, heat flow in rigid mechanical
structures happens mainly by conduction, which is gradient-driven but
has no curl contribution.  The impact of topology depends on the
various materials and is minimal for a perfectly homogeneous system.
In contrast, the curl component is very relevant for a fluid
(vorticity exists also when no net flow is present), and the topology
plays an important role too.  The gradient contribution for electrical
grids is the dominant factor, but curl and topology components are
also important.  The three components play a role in social systems,
where sometimes they may be difficult to interpret.

In particular, links do not represent a physical flow in social
systems.  Instead, they may represent correlation, influence,
proximity, probability of interaction, or inferred relationships,
depending on the nodes they connect.  Therefore, when computing the
curl one sums over quantities whose semantics are ambiguous.  This is
sharp contrast with the meaning of a non-zero curl in physics (local
rotation or current circulation, measurable) and in a traffic network
(circulation in a closed loop of the graph, consistent with a
conservation law and unambiguous).  How do we interpret an influence
loop $A \to B \to C \to A$ in human interactions?  Is it a feedback
loop (plausible), a cyclical behavior (maybe), inconsistency in data
(also possible!), a statistical artifact (very possible)?  There is no
unique interpretation.  Moreover, there is nothing like a conservation
law in social systems, contrary to the flow of physical systems.

Another delicate point is that many ``loops'' in social systems are
actually temporal cycles, not structural cycles.  Curl only models
instantaneous spatial loops, which may be nonexistent or irrelevant
compared to temporal dynamics.  Curl may be meaningful when there are
true feedback mechanisms, but its interpretation is context-dependent.
In other words, curl has a direct mechanistic interpretation grounded
in conservation laws in physical systems.  In data-driven systems, it
remains mathematically well-defined but its interpretation depends on
how interactions are modeled and may reflect feedback, inconsistency,
or statistical artifacts.  The issue is not mathematical; it is
semantic.

Can we find a unified way of modeling all the systems summarized in
table \ref{tab-classes}, arising from diverse domains?  Ideally, the
model should also be easy to interpret and be computationally
tractable.  This is the challenge addressed in this article.  We will
first address the mathematical aspects of continuous systems and then
focus on meaningful computational representations.  As we will see,
the recent Graph Vector Field (GVF) framework recently published by
\citet{GVF2026} provides a very general approach to achieve a unified
treatment of multimodal data, ensuring intrinsic interpretability and
explicit decomposition in the three components discussed above.  On
the other hand, its computational complexity and non-trivial topology
construction require careful attention on discretization choices.
Therefore, we will explore possible alternatives that, at the
potential price of reduced dynamic fidelity, possess advantages at
least as intermediate steps toward the full power of GVF.

The objective of this work is to investigate whether a unified and
computationally tractable modeling framework can be established for such
systems. We first review the mathematical foundations of flow decomposition
in continuous domains and examine their discrete counterparts. We then
analyze the GVF framework as a general and expressive
approach to modeling structured flows from data. Recognizing the practical
limitations of highly expressive models, we introduce a hierarchy of
reduced and alternative representations, and propose a cross-domain
validation strategy based on well-understood physical systems. This
approach enables a systematic exploration of the trade-offs between
model complexity, interpretability, and predictive performance.

The goal of this work is not to introduce a new mathematical
framework, but to position GVF within a broader hierarchy of
structured-flow models and to propose a methodology for model
selection, verification, and cross-domain validation.  This article is
intended as a conceptual and methodological contribution, establishing
the framework for future comparative studies.
In particular, it
positions GVF within the broader family of structured-flow models;
introduces a hierarchy of alternative model classes;
proposes GVF-guided model reduction and ablation strategies;
introduces a distinction between structural and behavioral verification;
and proposes a cross-domain validation methodology based on
well-understood physical systems.

\section{Continuous Flow Decomposition}

Continuous systems are often expressed as vector fields, possibly
obtained from suitable scalar potentials with the help of the gradient
($\nabla$) and curl ($\nabla\times$) differential operators.  Both of
them model local effects --- sources are connected to the divergence
$\nabla\cdot$, and rotations or vortices are connected to the curl
$\nabla\times$ --- but not the global structure.  The latter is
captured by the vector Laplacian operator
$\vec{\Delta} = \nabla(\nabla\cdot) - \nabla\times (\nabla\times)$ as
briefly summarized here.


Consider a $d$-dimensional real-valued vector field
\begin{equation}
\vec{F}(\vec{x},t) : \Omega \subset \mathbb{R}^d \rightarrow \mathbb{R}^d
\end{equation}
representing the flow of a conserved or transported quantity (e.g.,
mass, charge, energy) over a domain $\Omega$.
Typical governing equations relate the divergence and the curl of the
field to sources, sinks, and internal dynamics:
\begin{equation}
\nabla \cdot \vec{F} = \rho, \qquad \nabla \times \vec{F} = \vec{\omega},
\end{equation}
where the scalar $\rho$ represents a source (if $\rho>0$) or sink (if
$\rho<0$), and the vector $\vec{\omega}$ characterizes local
circulation.

If we consider a field $\vec{h}$ with zero divergence and zero curl,
then its Laplacian vector is clearly also zero:
\begin{equation}
\nabla \cdot \vec{h} = 0 \quad\text{and}\quad \nabla \times \vec{h} =  \vec{0} \quad\Rightarrow\quad \vec{\Delta} \vec{h} =  \vec{0},
\end{equation}
This field $\vec{h}$ only models non-local aspects.  In other words,
it captures the global features and the constraints imposed by the
topology of the space, on which the differential operators act
locally.
Because the components of the null vector are all zero, the previous
equation implies that each component $h_i$ of the field $\vec{h}$
separately satisfies Laplace's equation $\Delta h_i = 0$, where
$\Delta = \nabla\cdot\nabla$ is the scalar Laplace operator.  The
solutions of Laplace's equation are the harmonic functions, which
therefore model the global structure of the field.

Fields whose vector Laplacian is null cannot be expressed as
derivatives of potentials.  They encode degrees of freedom that are
not locally generated but globally constrained.  In domains with
non-trivial topology (e.g., containing holes or handles), such
harmonic fields correspond to flows that persist along independent
cycles.  Their dimension is determined by topological invariants
(e.g., Betti numbers), linking the existence of harmonic components
directly to the global structure of the domain \citep{Hatcher2002}.


When sufficiently regular, vector fields admit a decomposition in
terms of scalar and vector potentials.  The Helmholtz-Hodge
decomposition\footnote{The Helmholtz decomposition is the special case
  $\vec{h}=\vec{0}$ and is applicable in classical electrodynamics
  with the boundary condition that the fields decay at infinity.  Here
  we are interested into bounded domains, where the harmonic component
  is relevant.} states that, under appropriate boundary conditions,
\begin{equation}
 \label{eq-HHD} 
  \vec{F} = \nabla \phi + \nabla \times \vec{A} + \vec{h}
          = \vec{F}_{\text{grad}} + \vec{F}_{\text{curl}} + \vec{F}_{\text{harm}}
\end{equation}
where $\phi(\vec{x},t)$ is a scalar potential, $\vec{A}(\vec{x},t)$ is
a vector potential, and $\vec{h}(\vec{x},t)$ is a vector harmonic
field satisfying $\nabla \cdot \vec{h} = 0$ and
$\nabla \times \vec{h} = 0$.
This decomposition separates the flow into three independent
components:
\begin{itemize}
\item the \emph{gradient component}
  $\vec{F}_{\text{grad}}(\vec{x},t) =\nabla \phi$ is associated with
  sources and sinks, and is driven by local imbalances;
\item the \emph{curl component}
  $\vec{F}_{\text{curl}}(\vec{x},t) =\nabla \times \vec{A}$ is
  associated with local circulation, and represents cyclic or feedback
  dynamics;
\item the \emph{harmonic component}
  $\vec{F}_{\text{harm}}(\vec{x},t) =\vec{h}$ is constrained by the
  topology of the domain.
\end{itemize}

Note that we treat time as an external parameter $t$ governing the
evolution of the system.  At each time $t$, the field
$\mathbf{F}(x,t)$ admits a Helmholtz-Hodge decomposition into
gradient, curl, and harmonic components. The differential operators
involved act only on spatial variables, while the underlying
potentials and components are time-dependent.


The harmonic component $\vec{h}(\vec{x},t)$ is not determined by local
sources or circulation, but by the global topology of the domain
$\Omega$. In particular, its dimension is related to the number of
independent cycles (holes) in the domain \citep{Hatcher2002}.
In general, topology imposes constraints on admissible flows, defining
global pathways that cannot be removed by local modifications.

For example, consider a flow confined in a toroidal domain.  Even in
the absence of sources ($\rho = 0$) and local circulation
($\omega = 0$), a non-zero flow can persist along the toroidal
direction. Such a flow cannot be expressed as either a gradient or a
curl and is therefore captured by the harmonic component.

\begin{table}[h]
\caption{Continuous decomposition of flow fields.}
\label{tab-decomp}
  \centering
  \small
\begin{tabular}{lll}
\toprule
Component & Representation      & Interpretation \\
\midrule
Gradient  & $\nabla \phi$       & Source/sink \\
Curl      & $\nabla \times A$   & Loops, feedback \\
Harmonic  & kernel of Laplacian & Topology constraints \\
\bottomrule
\end{tabular}
\end{table}


This decomposition, summarized in Table \ref{tab-decomp}, provides a
natural framework for modeling flows across domains. However, its
practical application requires selecting a computational
representation of the domain and the vector field, which may involve
discretization or reduced-order modeling. The choice of representation
influences both the expressivity and the interpretability of the
resulting model.

\section{Computational Representations}

The continuous formulation described in the previous section provides
a general and domain-independent description of structured
flows. However, practical modeling, simulation, and inference require
a finite representation. This necessitates a transition from
continuous fields to discrete or reduced-order computational models.


Continuous systems governed by partial differential equations are
commonly discretized using numerical methods such as finite element
methods (FEM), finite volume methods, or finite difference schemes
\citep{Hughes1987,Bathe1995}.  In such approaches, the domain $\Omega$
is approximated by a mesh, and the vector field is represented by
values on nodes (or vertices) and links (or edges).

A key aspect is that this discretization is not unique: different mesh
choices may lead to different numerical properties and interpretations
of the same underlying system.  This is well known in FEM practice,
where both the placement of nodes and the choice of elements influence
accuracy, stability, and convergence.

From the perspective of flow decomposition, discretization also
determines how gradient, curl, and topology-constrained components are
represented.  In particular, the discrete analogs of divergence and
curl depend on the structure of the mesh.


An alternative mathematical representation consists in modeling the
system as a graph $G = (V, E)$, where vertices in the set $V$
represent spatial locations or entities, and edges in $E$ represent
admissible interactions or transport pathways.
This representation is natural for network-constrained systems, but it
may also be constructed from continuous systems via discretization or
from data via inference.  In graph-based models, flows are typically
represented on edges, and discrete analogs of gradient and divergence
operators can be defined using incidence matrices \citep{Lim2020}.

Graph representations offer computational efficiency and conceptual
simplicity, but they introduce modeling choices related to the
definition of nodes and links, which may affect interpretability.
Moreover, they only capture pairwise relations.


Graphs only make use of zero-dimensional (nodes) and one-dimensional
entities (links).  Including higher-dimensional elements like
triangles and tetrahedra, they can be extended to simplicial
complexes, which can be used to model higher-order interactions.
This also enables the definition of discrete exterior calculus
operators and a direct analog of the Helmholtz-Hodge decomposition in
a discrete setting \citep{Lim2020}.

Simplicial complex representations are particularly useful when
interactions cannot be fully captured by pairwise relations, and when
higher-order cycles play a role in the system dynamics.
Therefore, simplicial complexes offer a very attractive representation
for the modeling of complex systems, which is able to accommodate all
families summarized in Table \ref{tab-classes} above.

This is an important aspect, which motivated \citet{GVF2026} to
formulate an approach, the Graph Vector Field (GVF), capable of
addressing health-risk dynamics that cannot be fully captured by
mathematical graphs.
However, simplicial complexes may easily become computationally intractable.


In many applications, especially when computational resources are
limited or when interpretability is a priority, an attractive
alternative is provided by reduced-order models aiming to capture the
dominant components of the flow using a small number of parameters or
basis functions.
Examples include:
low-dimensional parametric representations,
linear dynamical systems,
reduced Hodge decompositions based on cycle bases,
probabilistic models with structured dependencies.

Compared to GVF, summarized in the next section, such approaches trade
expressive power for tractability and may serve as intermediate steps
toward more complex models, like the GVF approach.

Note that the choice of computational representation is not merely
technical, but has direct implications for
(1) the resolution at which the system is described,
(2) the topology of the resulting model, and
(3) the interpretation of gradient, curl, and harmonic components.

In particular, different discretizations or graph constructions may
lead to different topological structures, and therefore to different
decompositions of the same underlying continuous system. This
highlights the importance of carefully designing the mapping from data
or physical domains to computational representations.

\section{Graph Vector Field (GVF) Framework}

\citet{GVF2026} consider the problem of health risk assessment based
on digital data provided by heterogeneous sources. These include an
individual's medical history captured in electronic health records
(EHR), real-time physiological parameters monitored by wearable
devices (typically over long durations) or diagnostic medical devices
(typically over shorter periods), location information inferred from
smartphones, environmental data from meteorological infrastructure,
and genetic information. When available, spatiotemporal
epidemiological data describing disease incidence across geographical
regions may also be incorporated.

Such datasets are inherently complex, combining static
individual-specific information (e.g., genetic data) with dynamic
variables that evolve over time. The latter include longitudinal
measurements such as location and physiological signals, which are
often irregularly sampled, noisy, incomplete, or partially
unreliable. Individuals interact with their environment (e.g., weather
conditions, epidemiological context) and with other individuals (e.g.,
through co-location). However, the nature and strength of these
interactions are not directly observed in the data and must be
inferred from indirect evidence.
This indirect and partially latent structure poses significant
challenges for modeling and interpretation.

A key assumption in \citet{GVF2026} is that the complexity of such
systems cannot be adequately captured by simple graphs restricted to
pairwise interactions, and that higher-order structures, represented
by simplicial complexes, are required to model multi-way interactions
(e.g., triangles). A second requirement is the ability to represent
feedback loops, rather than only source-to-sink flows. Finally, the
authors emphasize the need to capture situations in which risk cannot
be mitigated by local interventions alone, but is instead constrained
by the global structure of the system.

To address these requirements, \citet{GVF2026} propose a framework
based on the discrete representation of vector fields and their
Helmholtz-Hodge decomposition \eqref{eq-HHD}, implemented via discrete
exterior calculus on simplicial complexes.  This approach, termed
Graph Vector Field (GVF), provides a unified representation of
gradient, curl, and topology-constrained components in a data-driven
setting.
This formulation naturally connects to classical vector field
decomposition, while extending it to data-driven and multimodal
settings.

\subsection{Mathematical Representation of the GVF}

The GVF framework provides a data-driven representation of complex
systems by modeling interactions as a time-dependent vector field
defined on a simplicial complex.  At each time $t$, the system is
represented by a simplicial complex $K(t)$, whose nodes correspond to
entities (e.g., individuals), and whose higher-order simplices encode
multi-way interactions inferred from the data.

A vector field $\vec{F}(t)$ is defined on the edges of $K(t)$,
representing the flow of a quantity of interest (e.g., health risk)
between connected entities.  This edge-based representation allows the
model to capture directional and distributed interactions across the
system.
The central idea of GVF is to decompose $\vec{F}(t)$ using a discrete
Helmholtz-Hodge decomposition:
\begin{equation}
  \label{eq-grad+curl+harm}
  \vec{F}(t) = \vec{F}_{\mathrm{grad}}(t)
             + \vec{F}_{\mathrm{curl}}(t)
             + \vec{F}_{\mathrm{harm}}(t),
\end{equation}
where:
\begin{itemize}
\item $\vec{F}_{\mathrm{grad}}(t)$ is a gradient component derived from a scalar
      potential defined on nodes, capturing source-to-sink behavior;
\item $\vec{F}_{\mathrm{curl}}(t)$ is a curl component defined on higher-order
      simplices, capturing local circulation and feedback loops;
\item $\vec{F}_{\mathrm{harm}}(t)$ is a harmonic component, lying in the kernel
      of the discrete Laplacian, capturing global constraints imposed by the
      topology of $K(t)$.
\end{itemize}

The decomposition is implemented using discrete exterior calculus,
which provides discrete analogs of gradient, divergence, and curl
operators on simplicial complexes. This allows the model to separate
local, cyclic, and global contributions to the flow in a
mathematically consistent way.

The basic ingredients for the discretization of a vector field
$\vec{F}$ are the mapping between continuous and discrete
representations --- 0-forms (scalars) become node values, 1-forms
(vectors) become edge values,
2-forms 
become face values --- and the representation of differential
operators as matrix multiplications:
\begin{eqnarray}
  \vec{F}_{\mathrm{grad}} & = & \mat{B}_{1}^{\top} \, \vec{\varphi}
  \label{eq-DEC-grad}
  \\
  \vec{F}_{\mathrm{curl}} & = & \mat{B}_{2} \, \vec{\psi}
  \label{eq-DEC-curl}
  \\
  \vec{F}_{\mathrm{harm}} & = & \mat{H} \, \vec{\alpha} 
  \label{eq-DEC-harm}
\end{eqnarray}
Similar to equation \eqref{eq-grad+curl+harm}, in which the vector
field is written as the sum of three components with the same
dimensionality, the discrete representation of the three components
above involves arrays with the same number $e$ of elements, equal to
the number of edges in the graph.
In equation \eqref{eq-DEC-grad}, $\mat{B}_{1}$ is the $(n \times e)$
node-edge incidence matrix, acting on the scalar field $\vec{\varphi}$
represented as an array of $n$ nodal values, and the result is an
array of $e$ edge values.
The matrix $\mat{B}_{2}$ is the $(e \times f)$ edge-face incidence
matrix, acting on the 2-form $\vec{\psi}$ represented as an array of
$f$ face values, and the result from equation \eqref{eq-DEC-curl} is
also an array of $e$ edge values.%
\footnote{The incidence matrices $\mat{B}_1$ and $\mat{B}_2$ encode
  only the topology and orientation of the simplicial complex. Their
  entries take values in $\{-1,0,+1\}$, reflecting the signed
  incidence between simplices. Metric information, if required, is
  introduced separately through weighting operators as shown in
  appendix \ref{sec-metric}.  }
Finally, also equation \eqref{eq-DEC-harm} results into an array of
$e$ edge values: $\mat{H}$ is an $(e \times b_1)$ matrix whose columns
span the harmonic subspace
$\ker L_1 = \ker \mat{B}_1 \cap \ker \mat{B}_2^{\top}$, which is the
kernel of the discrete 1-Hodge Laplacian
\begin{equation}
  L_1 = \mat{B}_1^{\top} \, \mat{B}_1 + \mat{B}_2 \, \mat{B}_2^{\top}
\end{equation}
The array of coefficients $\vec{\alpha}$ in equation
\eqref{eq-DEC-harm} has dimension $b_1$, the first Betti number of the
simplicial complex, which counts the number of independent cycles in
the simplicial complex, i.e., loops that are not boundaries of
higher-order simplices.%
\footnote{A loop is said to be ``filled'' if it bounds a 2-simplex (a
  face), in which case it does not contribute to the first Betti
  number.}

\subsection{Advantages and Limitations of GVF}

As summarized above, the mathematical representation of GVF provides a
unified treatment of multimodal data with intrinsic interpretability,
owing to the explicit decomposition into gradient, curl, and harmonic
components. This is particularly relevant to health risk modeling, as
each component may be associated with a different class of
risk-mitigation interventions \citep{GVF2026}. A gradient-dominated
flow may be attenuated by acting on sources, for example by reducing
environmental exposure. In contrast, risk sustained by feedback loops
may be mitigated by disrupting one or more links in the cycle, e.g.\
through targeted pharmacological interventions. When neither approach
is effective, mitigation may require modifying the underlying
interaction structure, for instance by introducing environmental or
social barriers.

In the implementation described by \citet{GVF2026}, the vector field
$\vec{F}(t)$ is learned from data through a parameterized model that
integrates heterogeneous inputs (e.g., physiological, environmental,
and contextual data). The resulting decomposition yields interpretable
components associated with different mechanisms driving the system
dynamics, which constitutes a significant advantage.

From a mathematical perspective, GVF provides a natural and expressive
framework for representing complex dynamical systems. From a
scientific standpoint, however, its effectiveness depends on the
validity of the underlying hypothesis, namely that health-risk
dynamics can be meaningfully represented as an edge flow on a graph or
simplicial complex. This assumption is context- and
application-dependent, and does not constitute a general property of
the method itself, which remains well-founded mathematically.
Indeed, the necessity of simplicial representations is itself an
empirical question and should be assessed through ablation studies, as
illustrated in section \ref{sec-model-ana} below.

When predictive performance is a primary objective, GVF may be less
attractive than simpler models that focus on dominant effects rather
than capturing all components of the dynamics. Such reduced models may
achieve better generalization, require less data, and be easier to
validate.

From an implementation perspective, GVF entails combinatorial
complexity, a strong dependence on discretization choices, and
non-trivial topology construction. These aspects can significantly
impact development effort and computational cost.  Moreover, the large
number of degrees of freedom in a full GVF model may require
substantially larger datasets than lower-dimensional alternatives for
reliable training.\footnote{The Helmholtz--Hodge decomposition
  provides an optimal representation of a given flow field, but does
  not guarantee that such a field can be reliably inferred from noisy,
  heterogeneous data.}  This may represent a practical limitation for
the application of this otherwise elegant and interpretable framework.

This observation motivates the exploration of reduced-order and
alternative modeling approaches, discussed in the following section.

\section{Alternative Modeling Approaches}

We consider here a set of alternative modeling approaches to GVF that
are computationally simpler and, depending on the application, may be
sufficient to capture the dominant effects. This is not intended as a
criticism of the GVF framework, which is mathematically well-founded
and offers both interpretability and the ability to learn its
components from data, but rather reflects a pragmatic perspective. In
many practical settings --- such as modeling the spread of an
infectious disease to produce timely forecasts and actionable insights
--- computational efficiency and data requirements are critical
constraints. In such cases, a simpler model that yields a timely and
robust approximation may be preferable to a more expressive approach
whose estimation and validation require substantial resources.

\subsection{Parametric Conditional Models}

A possible approach to modeling structured flows, often used to model
complex physical systems, consists in specifying a low-dimensional
parametric representation in which the relationships between variables
are explicitly encoded, rather than inferred from data.  In this
framework, the system is described by a set of variables whose
interactions are modeled through parametric functions, possibly with
hierarchical or conditional dependencies.  These relationships are
defined based on prior knowledge of the system, implicitly assuming
that the dominant mechanisms governing the dynamics can be captured by
a limited number of parametric dependencies.

In contrast to GVF, which seeks to infer both the structure of
interactions and the associated flow field from data, parametric
conditional models assume that the main mechanisms governing the
system are known a priori and can be expressed through a small number
of parameters.  For instance, one may write the evolution of a
$d$-dimensional state variable $\vec{x}(t)$ as
\begin{equation}
  \vec{x}(t+1) = f\big(\vec{x}(t), \vec{\theta}(\vec{x}(t))\big),
\end{equation}
where the parameters encoded in the $k$-dimensional vector
$\vec{\theta}$ may themselves depend on other variables, capturing
conditional dependencies in a structured way.

From the perspective of flow decomposition, the gradient, curl, and
topology-constrained components are not derived from a learned field
on a discrete structure, but are instead encoded implicitly in the
functional form of the model. For example, source-driven effects may
be represented through potential-like terms, feedback mechanisms
through cyclic dependencies between variables, and global constraints
through explicit coupling terms.

This approach leads to a fundamentally different modeling
philosophy. Rather than learning a high-dimensional representation of
the system, including its topology, from data, one imposes a
simplified structure that captures the dominant mechanisms. As a
consequence, the number of degrees of freedom to be estimated is
significantly reduced.

The reduction in model complexity has important practical implications. In
particular, parametric conditional models 
(1) require substantially less training data,
(2) are less sensitive to noise and missing observations,
(3) are easier to interpret and validate, and
(4) can be evaluated and calibrated with limited computational resources.

These advantages come at the cost of reduced expressive power. By
construction, the model may not capture fine-grained interactions or
emergent structures that are not explicitly encoded. Therefore, the
suitability of this approach depends on the extent to which the
dominant dynamics of the system can be described by a small number of
well-chosen parameters.

In settings where data are scarce, noisy, or partially observed, and
where the underlying mechanisms are reasonably well understood,
parametric conditional models provide a viable and often effective
alternative to more expressive data-driven approaches such as GVF.

The necessary condition for this approach is that the dominant
mechanisms driving the system dynamics are known. However, this
assumption is often not satisfied for many of the systems listed in
Table~\ref{tab-classes}. As will be discussed in the following, a
possible heuristic strategy to uncover these mechanisms is to use GVF
as a diagnostic tool, by systematically assessing how well the
observed data can be explained by gradient, curl, and
topology-constrained components.

\subsection{Linear Graph Dynamical Models}

A complementary approach consists in representing the system dynamics
through linear evolution models defined on a graph. In this framework,
the state of the system is represented by a vector
$\vec{x}(t) \in \mathbb{R}^n$, whose components correspond to
variables associated with the nodes of a graph. The temporal evolution
may be for example implemented as a Markov process, with a system of
linear equations connecting the current to the previous state:
\begin{equation}
  \label{eq-LGDM}
  \vec{x}(t+1) = \mat{A} \, \vec{x}(t) + \vec{u}(t)
\end{equation}
Here, $\mat{A} \in \mathbb{R}^{n \times n}$ is a matrix encoding
interactions between nodes, and $\vec{u}(t)$ represents external
inputs or forcing terms.\footnote{ When time evolution is derived from
  a continuity equation, \eqref{eq-LGDM} takes the form
  \eqref{eq-time-evol} from section \ref{sec-time-evol} below.
  Details are provided in appendix \ref{app-dynamic}.  }

The matrix $\mat{A}$ may be constrained by the graph structure, for
example by enforcing that $\mat{A}_{ij} \neq 0$ only if there exists
an edge between nodes $i$ and $j$.  Additional structure may be
imposed, such as symmetry, sparsity, or conservation constraints,
depending on the application. In many cases, $\mat{A}$ is related to
graph operators such as the adjacency matrix or the graph Laplacian
\citep{Lim2020}.

From the perspective of flow modeling, linear graph dynamical systems
provide an implicit representation of transport and interaction
processes. Gradient-like effects correspond to diffusive or
dissipative terms, often modeled through Laplacian operators, while
cyclic or feedback behavior may arise from asymmetric or non-normal
components of $\mat{A}$. However, unlike GVF, these components are not
explicitly separated, but are instead embedded in the structure of the
evolution operator. The limitations of such linear representations in
capturing complex dynamics and higher-order interactions are also
discussed in \citet{GVF2026}.

Compared to parametric conditional models, linear graph dynamical
models adopt a different philosophy. While parametric models
explicitly encode the functional form of interactions based on prior
knowledge, linear graph models infer a global interaction structure
directly from data through the matrix $\mat{A}$, typically without
imposing strong prior constraints beyond sparsity or topology.  In
other words, whereas parametric models attribute specific roles to
individual mechanisms, linear graph models capture aggregate
interactions without explicitly separating their underlying causes.
As a result, they offer greater flexibility than fully specified
parametric models, but at the cost of reduced interpretability of
individual mechanisms.

From a practical perspective, linear graph models provide a
significant reduction in complexity compared to GVF. The dynamics are
governed by a single matrix, leading to a relatively small number of
parameters and enabling efficient estimation from data. Furthermore,
linear systems are well understood, allowing for stability analysis,
spectral characterization, and, in some cases, closed-form solutions.

The main limitation of this approach is its restricted expressive
power. Linear models may fail to capture nonlinear interactions,
higher-order dependencies, and explicit topological constraints that
are naturally represented in GVF. As a result, they are best suited
for systems in which the dominant behavior can be approximated by
linear interactions around an operating point.

Despite these limitations, linear graph dynamical models provide a
useful intermediate representation between fully data-driven
approaches such as GVF and highly structured parametric models. They
offer a favorable trade-off between expressivity, interpretability,
and computational efficiency, particularly in settings where data are
limited and rapid inference is required.

\subsection{Reduced Hodge and Cycle-Basis Models}\label{sec-HHD-reduced}

A natural intermediate approach between fully expressive frameworks
such as GVF and simpler linear or parametric models consists in
retaining the structural decomposition of flows while reducing the
dimensionality of the representation.  This can be achieved by
restricting the Helmholtz-Hodge decomposition to a low-dimensional
basis of relevant components.

In the discrete setting, the decomposition of an edge-based field
$\vec{F} \in \mathbb{R}^e$ can be written as
\begin{equation}
  \label{eq-HDD}
  \vec{F}(t) = \mat{B}_1^{\top} \, \vec{\varphi}(t) + \mat{B}_2 \, \vec{\psi}(t) + \mat{H} \, \vec{\alpha}(t),
\end{equation}
where $\mat{B}_1$ and $\mat{B}_2$ are incidence matrices, and
$\mat{H}$ spans the harmonic subspace of dimension $b_1$. While this
representation is exact, the dimensions of $\vec{\varphi}$,
$\vec{\psi}$, and $\vec{\alpha}$ scale with the number of nodes,
faces, and independent cycles, respectively, which may become
prohibitive in large systems.

Here we present a procedure to decrease the dimensionality of this
model, based on singular value decomposition (SVD).  The procedure
first performs a change of basis, then discards the basis elements not
associated with strong enough patterns

Starting from the full discrete Helmholtz-Hodge decomposition
\eqref{eq-HDD} we define the three component fields
\begin{equation}
  \vec{F}_{\text{g}}(t) = \mat{B}_1^{\top}\,\vec{\varphi}(t),\qquad
  \vec{F}_{\text{c}}(t) = \mat{B}_2\,\vec{\psi}(t),\qquad
  \vec{F}_{\text{h}}(t) = \mat{H}\,\vec{\alpha}(t).
\end{equation}
A reduced representation can be obtained by collecting
\emph{representative} snapshots\footnote{No need to take all available
  data. It is necessary and sufficient to take a subset that captures
  the relevant dynamics. The quality of the reduced model depends more
  on the representativeness of the chosen dataset than on its size.}
of each component and applying a singular value decomposition
separately:
\begin{equation}
  \label{eq-svd}
  \begin{split}
    \mat{X}_{\text{g}} &=
    \begin{bmatrix}
      \vec{F}_{\text{g}}(t_1) & \vec{F}_{\text{g}}(t_2) & \cdots
    \end{bmatrix}\;
                = \mat{U}_{\text{g}} \, \mat{\Sigma}_{\text{g}} \, \mat{V}_{\text{g}}^{\top},
\\
    \mat{X}_{\text{c}} &=
    \begin{bmatrix}
      \vec{F}_{\text{c}}(t_1) & \vec{F}_{\text{c}}(t_2) & \cdots
    \end{bmatrix}\;\,
                = \mat{U}_{\text{c}} \, \mat{\Sigma}_{\text{c}} \, \mat{V}_{\text{c}}^{\top},
\\
    \mat{X}_{\text{h}} &=
    \begin{bmatrix}
      \vec{F}_{\text{h}}(t_1) & \vec{F}_{\text{h}}(t_2) & \cdots
    \end{bmatrix}
                = \mat{U}_{\text{h}} \, \mat{\Sigma}_{\text{h}} \, \mat{V}_{\text{h}}^{\top}.
  \end{split}
\end{equation}
The columns of $\mat{U}_{\text{g}}$, $\mat{U}_{\text{c}}$, and
$\mat{U}_{\text{h}}$ define orthonormal bases for the dominant
gradient, curl, and harmonic flow patterns observed in the training
data.
Each column of $\mat{U}_{\text{g}}$ is an orthonormal basis vector
spanning the space of edge flows, ordered by decreasing contribution
to the variance of the data, and represents a recurring structure in
the data.
The matrices
$\mat{\Sigma}_{\text{g}} = \text{diag}(\sigma_{\text{g},1}, \sigma_{\text{g},2}, \ldots)$,
$\mat{\Sigma}_{\text{c}} = \text{diag}(\sigma_{\text{c},1}, \sigma_{\text{c},2}, \ldots)$, and
$\mat{\Sigma}_{\text{h}} = \text{diag}(\sigma_{\text{h},1}, \sigma_{\text{h},2}, \ldots)$ are diagonal,
and their elements are the singular values of $\mat{X}_{\text{g}}$,
$\mat{X}_{\text{c}}$, and $\mat{X}_{\text{h}}$, respectively.
Finally, the rows of the matrices $\mat{V}_{\text{g}}$,
$\mat{V}_{\text{c}}$, and $\mat{V}_{\text{h}}$ are indexed by the
snapshot time index $j$, which correspond to the time samples $t_j$.
For example, from the first equation in \eqref{eq-svd} one obtains
\(
  \vec{F}_{\text{g}}(t_j) = \sum_{k=1}^{r_{\text{g}}} \mat{U}_{\text{g},k} \, \sigma_{\text{g},k}  \mat{V}_{\text{g},jk} 
\),
where $r_{\text{g}}$ is the rank.  

Now the gradient component can be written in terms of a new basis:
\begin{equation}
  \vec{F}_{\text{g}} = \mat{B}_1^{\top} \, \vec{\varphi} = \mat{U}_{\text{g}} \, \vec{a}_{\text{g}}
  \qquad\text{with}\qquad
  \vec{a}_{\text{g}} = \mat{U}_{\text{g}}^{\top} \, \mat{B}_1^{\top} \, \vec{\varphi}
  = (\mat{B}_1 \, \mat{U}_{\text{g}})^{\top} \, \vec{\varphi}
\end{equation}
with similar expressions for the curl and harmonic components.  At
this stage, we are still considering the full (non-truncated) SVD
representation, in which the decomposition is exact.  Our next goal is
to identify the most important basis vectors and ignore the others, to
reduce the model dimensionality.

Singular values are real numbers sorted in decreasing order
($\sigma_{\text{g},1} \ge \sigma_{\text{g},2} \ge \ldots \ge 0$
reflecting the sampled data) and, to reduce the model dimensionality,
we select the smallest set of $K_{\text{g}}$ such that the cumulative
singular-value energy exceeds $\eta_{\text{g}}=0.9$ (a different
threshold may also be used):
\begin{equation}
  \frac{\sum_{k=1}^{K_{\text{g}}}\sigma_{\text{g},k}^2}
       {\sum_{k}\sigma_{\text{g},k}^2} \ge \eta_{\text{g}}
\end{equation}
We proceed the same way also for the curl and harmonic components.
Keeping only the first $K_{\text{g}}$, $K_{\text{c}}$, and
$K_{\text{h}}$ columns of $\mat{U}_{\text{g}}$, $\mat{U}_{\text{c}}$,
and $\mat{U}_{\text{h}}$, respectively, gives the reduced model
\begin{equation}
  \label{eq-HHD-reduced}
  \vec{F}(t) \approx \mat{U}_{\text{g,red}}\, \vec{a}_{\text{g}}(t)
                   + \mat{U}_{\text{c,red}}\, \vec{a}_{\text{c}}(t)
                   + \mat{U}_{\text{h,red}}\, \vec{a}_{\text{h}}(t),
\end{equation}
where $\vec{a}_{\text{g}}(t)$, $\vec{a}_{\text{c}}(t)$, and
$\vec{a}_{\text{h}}(t)$ are the reduced components.
The thresholds $\eta_{\text{g}},\eta_{\text{c}},\eta_{\text{h}}$ may
be then varied to assess the sensitivity of the results to model
reduction.
After truncation, the representation \eqref{eq-HHD-reduced} becomes
approximate.

This provides a practical reduction procedure. First, the full model
is run on controlled or representative data and the three decomposed
components are recorded. Second, separate SVDs are performed on the
gradient, curl, and harmonic snapshot matrices. Third, the dominant
modes are retained according to a chosen energy threshold. Finally,
the reduced model is trained and evaluated for different
thresholds. If lowering the threshold does not significantly affect
the prediction or decomposition metrics, the reduced basis is
considered sufficient; otherwise, additional modes are retained.

This approach avoids learning the full high-dimensional field
directly.  Instead, it represents the field in a reduced basis learned
from data, significantly reducing the number of degrees of freedom.
The reduced basis provides a data-driven approximation of the image of
the discrete differential operators, capturing the dominant flow
patterns actually observed in the representative subset of data chosen
for the SVD.

From a modeling perspective, reduced Hodge representations retain the
key interpretability advantages of the full decomposition, as each
component is still associated with a specific physical or structural
mechanism (source/sink, circulation, topology). However, they do not
aim to reconstruct all fine-grained details of the flow, and therefore
may miss localized or higher-order effects.

Overall, reduced Hodge and cycle-basis models provide a principled
compromise between expressivity and tractability. They preserve the
conceptual clarity of the Helmholtz-Hodge decomposition while enabling
practical implementation in settings where data are limited and
computational resources are constrained.
In particular, this approach enables systematic exploration of model
robustness through repeated evaluation on synthetic datasets, which is
often impractical for higher-dimensional models.

\subsection{Time Evolution}\label{sec-time-evol}

The GVF framework, as introduced above, provides a decomposition of
the vector field $\vec{F}(t)$ at a given time $t$, but does not by
itself define the temporal evolution of the system. In contrast to
parametric conditional models and linear graph dynamical systems,
which explicitly specify an evolution rule linking states at
successive time steps, GVF is inherently an instantaneous
representation.
The same remark applies to reduced Hodge models, which inherit the
decomposition structure of GVF but do not, in general, specify the
time evolution unless an explicit dynamical law is introduced.

More precisely, temporal loops that take a shorter time than the
sampling time $\Delta t$ leave their ``imprint'' on the GVF topology,
while those lasting longer than the sampling time $\Delta t$ are not
captured by the underlying geometrical structure.  They need to be
described by an explicit time evolution relationship.

For example, a dynamical model consistent with a continuous physical
system is obtained by coupling the flow field with a conservation
equation describing the local trasport of some quantity.  Its
differential form states that the time derivative of the local density
\vec{x} equals the sum of the source term \vec{s} --- which represents
a sink if negative --- and of the net inflow of that quantity, given
by minus the divergence of its current:
\begin{equation}
  \label{eq-continuity}
  \frac{\partial \vec{x}}{\partial t} = \vec{s} - \nabla \cdot \vec{J}
\end{equation}
(the current \vec{J} models the local outward flow).

The discretized version of the inflow $-\nabla\cdot\vec{J}$ at each
node is the matrix product $\mat{B}_1\,\vec{F}$, where $\mat{B}_1$ is
the same node-edge incidence matrix featuring in equation
\eqref{eq-DEC-grad} above.\footnote{The transposed $\mat{B}_1$ matrix
  is used for the discrete gradient in equation \eqref{eq-DEC-grad}.}
Its elements\footnote{For the pure topology.  Metric needs
  introduction of weights as shown in appendix \ref{sec-metric}.}  are
$(\mat{B}_1)_{ve} = +1$ if edge $e$ enters node $v$; $-1$ if edge $e$
leaves node; or $0$ otherwise.
Therefore, the continuity equation governing the time evolution of the
model can be written in a discrete form as
\begin{equation}
  \label{eq-time-evol}
  \vec{x}(t+1) = \vec{x}(t) + \Delta t \, [ \vec{s}(t) + \mat{B}_1\,\vec{F}(t) ]
\end{equation}
where \vec{s} is an external source.

In GVF the field $\vec{F}(t)$ is itself decomposed into gradient,
curl, and harmonic components.  This provides a structured
representation of the driving mechanisms, while the evolution equation
above governs the temporal dynamics.
This separation between spatial structure and temporal evolution
allows the modeler to independently design the dynamical law and the
decomposition of underlying mechanisms.
The decomposition and the evolution law are conceptually distinct.
Different applications may require different conservation or evolution
equations.

\subsection{Model Analysis}\label{sec-model-ana}

Compared to GVF, the parametric conditional model described above
represents a fundamentally different approach, as it does not rely on
an explicit topological representation inferred from data. In
contrast, linear graph models and reduced Hodge representations may be
viewed as constrained or simplified instances of the GVF framework.

In particular, GVF can be interpreted as a general model from which
simpler representations may be derived by restricting the functional
space or the underlying topology. For example, linear graph dynamical
models correspond to a representation in which interactions are
encoded in a single operator without explicit decomposition, while
reduced Hodge models retain the decomposition but restrict it to a
low-dimensional basis.

\begin{table}[t]
\caption{Comparison of modeling approaches.}
  \centering
  \small
\begin{tabular}{llllllll}
\toprule
Model & Complexity & Interpretability & Expressivity & Dimensionality\\
\midrule
Parametric & low      & high   & medium      & very low \\
Linear     & very low & medium & low         & low \\
Reduced    & medium   & high   & medium-high & medium \\
GVF        & high     & high   & high        & high \\
\bottomrule
\end{tabular}
\end{table}

\begin{table}[t!]
\caption{Comparison of modeling approaches (continued).}
  \centering
  \small
\begin{tabular}{llllllll}
\toprule
Model & Structure & Topology & Learning \\
\midrule
Parametric & imposed a priori     & fixed, implicit   & parameters \\
Linear     & learned globally     & fixed             & matrix \\
Reduced    & learned + decomposed & explicit, reduced & coefficients \\
GVF        & learned + decomposed & explicit, learned & whole field \\
\bottomrule
\end{tabular}
\end{table}

This observation suggests a practical strategy for model
analysis. Starting from a full GVF implementation, one may selectively
restrict or suppress individual components of the decomposition to
assess their relative contribution. For instance, setting the curl
component to zero removes cyclic interactions, yielding a purely
gradient-driven model, while removing the gradient component
emphasizes circulation and topology-driven effects. Similarly,
restricting the harmonic component suppresses contributions associated
with global topological constraints.  Appendix \ref{sec-unified}
provides a unified notation that supports this analysis.

In addition to component-wise analysis, the complexity of the
underlying topological representation may be varied. For example, one
may reduce the number of higher-order simplices to limit the space of
admissible circulations, or simplify the graph structure to focus on
dominant interactions. Such controlled simplifications provide a
systematic way to evaluate the sensitivity of the model to topological
assumptions and to identify the minimal structure required to capture
the observed dynamics.

Overall, this hierarchical perspective enables the use of GVF not only
as a predictive model, but also as a \emph{diagnostic tool} for
understanding the relative importance of gradient, curl, and
topology-constrained components in complex systems.
This approach is analogous to ablation studies in machine learning,
where the impact of individual components is evaluated by selectively
disabling them.

\begin{figure}[t!]
\centering
\resizebox{\textwidth}{!}{%
\begin{tikzpicture}[
    font=\small,
    box/.style={
        draw,
        rounded corners,
        align=center,
        minimum width=3.7cm,
        minimum height=1.15cm,
        fill=blue!4
    },
    widebox/.style={
        draw,
        rounded corners,
        align=center,
        minimum width=6.3cm,
        minimum height=1.45cm,
        fill=blue!4
    },
    actionbox/.style={
        draw,
        rounded corners,
        align=center,
        minimum width=6.2cm,
        minimum height=1.35cm,
        fill=orange!8
    },
    arrow/.style={
        -{Latex[length=2.5mm]},
        thick
    },
    dashedarrow/.style={
        -{Latex[length=2.5mm]},
        thick,
        dashed
    },
    note/.style={
        align=center,
        font=\scriptsize
    }
]

\node[box] (gvf) at (0,0)
{Full GVF\\[-1mm]
{\scriptsize maximum expressivity}};

\node[box, right=-0.1cm of gvf, yshift=-1.5cm] (hodge)
{Reduced Hodge\\[-1mm]
{\scriptsize truncated decomposition}};

\node[box, right=-0.1cm of hodge, yshift=-1.5cm] (linear)
{Linear Graph Model\\[-1mm]
{\scriptsize single evolution operator}};

\node[box, right=-0.1cm of linear, yshift=-1.5cm] (param)
{Parametric Conditional\\[-1mm]
{\scriptsize mechanisms encoded a priori}};

\draw[arrow] (gvf.south) -- node[below left,note]{restrict\\basis} (hodge.north west);
\draw[arrow] (hodge.south) -- node[below left,note]{drop / merge\\components} (linear.north west);
\draw[arrow] (linear.south) -- node[below left,note]{define\\coupling} (param.north west);

\node[font=\scriptsize\itshape, anchor=east] at ($(gvf.north west)+(+2.8,+0.7)$)
{higher expressivity};
\node[font=\scriptsize\itshape, anchor=west] at ($(param.north east)+(-2.5,+1.2)$)
{lower complexity};

\draw[arrow] ($(gvf.north west)+(2.8,+0.7)$) --
             ($(param.north east)+(-2.5,1.2)$);

\node[widebox, fill=green!4, 
      below=5.0cm of gvf, xshift=2.1cm,
      minimum width=6.0cm, minimum height=2.7cm] (ablation)
{\textbf{Component-wise ablation}\\[1mm]
{\scriptsize Starting from the full GVF,}\\[-1mm]
{\scriptsize selectively suppress components}\\[2mm]
\begin{tabular}{cccc}
{\scriptsize no gradient} &
{\scriptsize no curl} &
{\scriptsize no harmonic} &
{\scriptsize gradient only} \\[-1mm]
{\scriptsize (sources off)} &
{\scriptsize (loops off)} &
{\scriptsize (topology off)} &
{\scriptsize (reduced field)}
\end{tabular}};

\node[widebox, fill=purple!4, below=2.0cm of linear, xshift=3.1cm,
      minimum width=5.5cm, minimum height=2.7cm] (topology)
{\textbf{Topology simplification}\\[1mm]
{\scriptsize reduce graph complexity}\\
{\scriptsize remove selected faces}\\
{\scriptsize retain only dominant cycles}};

\draw[dashedarrow] (ablation.east) --
node[above,note]{analyze\\ sensitivity}
(topology.west);

\node[actionbox, fill=green!5, below=1.65cm of ablation,
      minimum width=7.7cm] (mechanism)
{\textbf{Dominant mechanisms identified}\\
{\scriptsize determine which components and structures}\\[-1mm]
{\scriptsize drive the observed dynamics}};

\node[actionbox, fill=orange!8, below=1.65cm of topology,
      minimum width=5.5cm] (custom)
{\textbf{Custom parametric model}\\
{\scriptsize encode identified mechanisms}\\
{\scriptsize with few trainable parameters}};

\draw[arrow] (ablation.south) -- (mechanism.north);
\draw[arrow] (topology.south) -- (custom.north);
\draw[dashedarrow] (mechanism.east) --
node[above,note]{model\\reduction}
(custom.west);

\node[below=1.0cm of mechanism, xshift=-2.8cm, note] (data)
{\normalsize Data};

\node[right=1.5cm of data, note] (model)
{\normalsize Model\\{\footnotesize(choose complexity)}};

\node[right=1.5cm of model, note] (pred)
{\normalsize Prediction\\ \normalsize and evaluation};

\node[right=1.5cm of pred, note] (action)
{\normalsize Action /\\ \normalsize intervention};

\draw[arrow] (data) -- (model);
\draw[arrow] (model) -- (pred);
\draw[arrow] (pred) -- (action);

\draw[dashedarrow]
(action.south) .. controls +(0,-1.0) and +(0,-1.0) .. (data.south);

\node[font=\footnotesize\itshape, align=center, xshift=-2.2cm,
      below=0.65cm of pred]
{Iterative modeling--prediction--intervention loop};

\end{tikzpicture}%
}
\caption{Hierarchy of reduced models and ablation-based analysis
  starting from a full GVF representation.  Restricting the
  decomposition basis, simplifying the topology, or suppressing
  individual components can reveal dominant mechanisms and guide the
  construction of compact parametric models.}
\label{fig:gvf_model_reduction}
\end{figure}
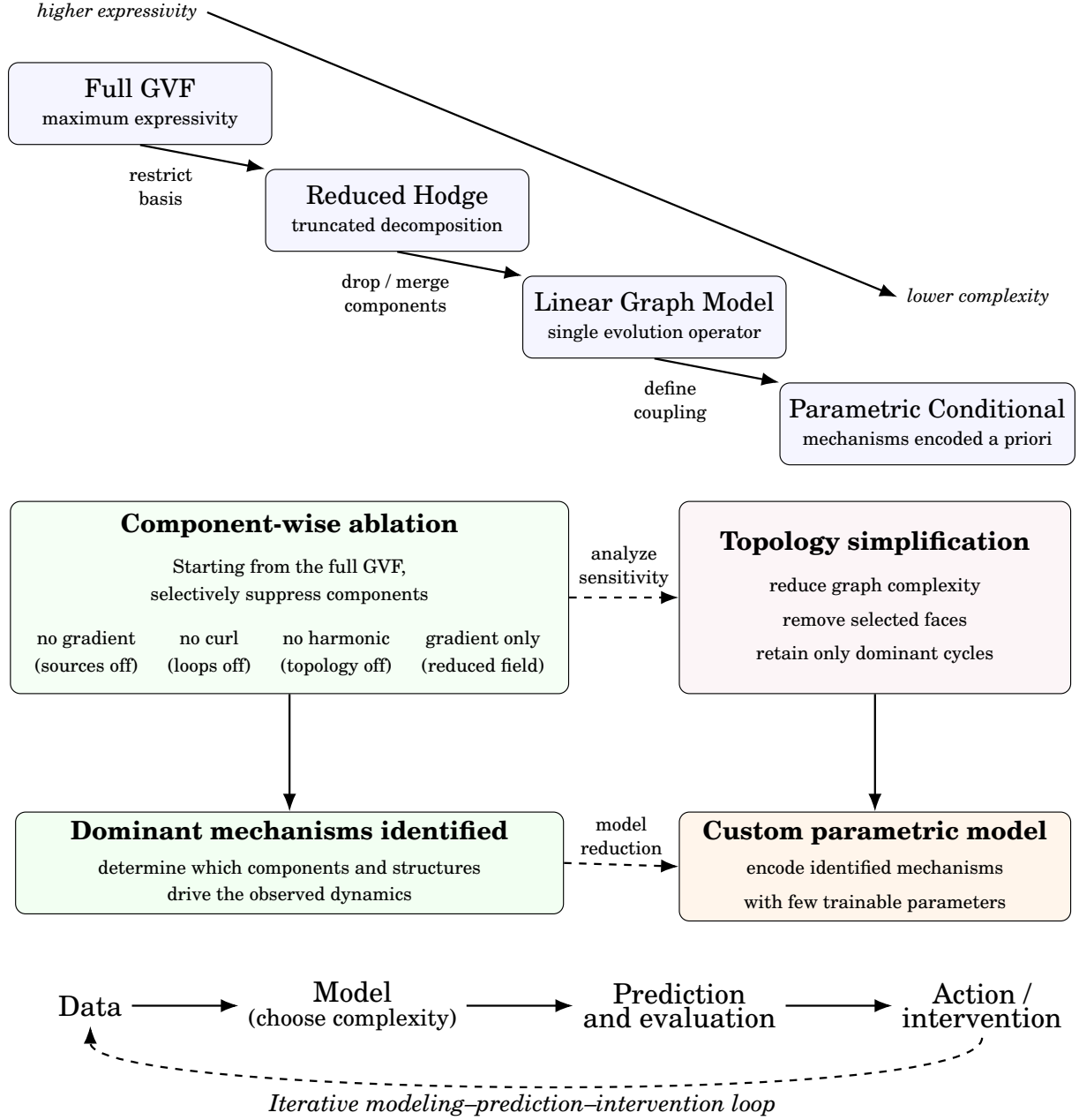

This ablation-based exploration can also be used as a tool for
hypothesis generation.
In other words, GVF can be used as a scientific instrument for model discovery.
By systematically restricting GVF components and topology, one may
identify dominant mechanisms that govern the observed dynamics. Such
insights can guide the design of custom parametric models that encode
these mechanisms explicitly, potentially improving predictive
performance while reducing model complexity. This strategy may be
particularly valuable in applications where the iterative
modeling--prediction--intervention loop must operate under tight time
and data constraints.  Figure \ref{fig:gvf_model_reduction}
schematically represents how this task may be iteratively approaced.

\section{Model Validation}

The performance of a predictive model depends on several key factors,
among which the mathematical formulation, the implementation, and the
training data play central roles. The previous sections have addressed
the mathematical aspects, highlighting that the GVF framework provides
a general and expressive representation capable, in principle, of
adapting to a wide variety of systems.  It was also noted that such
expressivity may exceed practical requirements, and that reduced or
alternative models may be preferable depending on the application. In
this context, GVF itself may be used as an exploratory tool to
identify the appropriate level of model complexity. For the remainder
of this section, we therefore assume that the mathematical formulation
is adequate.

The next critical aspect is the correctness of the
implementation. Before expecting meaningful insights or reliable
predictions, it is necessary to verify that the model has been
implemented without flaws. A standard approach consists in evaluating
the model on datasets generated from systems whose behavior is well
understood. In such cases, the expected outcomes are known, and the
ability of the model to recover key features provides a direct
validation of the implementation.

Before entering into details, it is worthwhile to clarify the notion
of ``correctness'' for data-driven models. In systems engineering, a
distinction is made between \emph{verification}, which concerns
compliance with specifications, and \emph{validation}, which concerns
the fitness for the intended purpose.  Verification addresses the
question ``Did we build the system right?'', whereas validation asks
``Did we build the right system?''
The terms verification and validation are used here in the
systems-engineering sense.

In the present context, which involves models whose parameters are
inferred from data, this distinction requires some
adaptation. Verification applies naturally to the mathematical
formulation and to its implementation. However, the resulting system
operates with parameters learned from data, and these parameters are
not subject to explicit requirements; their correctness can only be
assessed in statistical terms.

This motivates a further distinction between \emph{structural
  verification} and \emph{behavioral verification}. Structural
verification ensures that the model is implemented correctly,
including the consistency of equations, operators, and
decomposition. Behavioral verification, on the other hand, evaluates
whether the model exhibits the expected behavior when applied to
controlled datasets with known ground truth. In particular, when
trained on data generated from simple, well-understood systems, the
model should correctly identify and reconstruct the relevant
components, such as gradient, curl, and harmonic contributions.

Only after these verification steps have been successfully completed
can the model be applied to the system of interest. Even then, full
validation may remain out of reach, due to the inherent limitations of
real-world data, which may be noisy, biased, incomplete, or scarce,
and to possible mismatches between model assumptions and
reality. Nevertheless, verification provides an important guarantee:
if the model fails to produce meaningful results on real data, the
most likely cause lies in the data rather than in the formulation or
implementation of the model.

This raises a central question to the practical deployment of
data-driven models: what constitutes an appropriate training dataset
for model verification?

Ideally, such verification datasets should be derived from systems for
which the underlying dynamics are fully specified, either analytically
or through high-fidelity simulation, and for which sufficient data can
be generated. In practice, this condition is not always met in the
target application domain.  Therefore, it is essential to identify
alternative systems and datasets that can serve as proxies for
validation purposes.

Once both the mathematical formulation and the implementation have
been validated against controlled datasets, the model can be applied
to the system of interest. If, at this stage, the model fails to
provide relevant insights or accurate predictions, the most likely
cause lies in the limitations of the training data, rather than in the
formulation or implementation of the model.

Two common issues may arise: insufficient data quantity or
insufficient data quality. In the case of data scarcity, a practical
strategy consists in adopting a simpler model with fewer degrees of
freedom, thereby reducing the amount of data required for reliable
estimation. The expected loss in fidelity can be quantified in advance
by evaluating the simplified model on controlled datasets, using the
same ablation-based methodology described in the previous section.
This provides an a priori assessment of the trade-off between model
complexity and predictive accuracy.

In contrast, limited data quality represents a more fundamental
challenge, as it cannot be easily compensated by model
simplification. Improving data quality typically requires significant
effort in data acquisition, preprocessing, and validation. For this
reason, it is crucial to establish the correctness of the model using
independent and well-characterized datasets before investing resources
in improving the quality of the target data.

The central objective is therefore to identify suitable systems and
datasets for model verification. Systems governed by well-established
physical laws are particularly attractive in this respect, as they
provide a clear ground truth and often allow for the generation of
large, high-quality datasets through simulation. When such datasets
are publicly available, they offer a practical and efficient means to
validate both the implementation and the behavior of the model under
controlled conditions.

This separation between model validity and data limitations is
essential to avoid misattributing model failure to incorrect causes.

\section{Verification Datasets and Cross-Domain Applicability}

The verification strategy outlined above requires datasets generated
from systems whose governing mechanisms are sufficiently well
understood. The ideal dataset should satisfy the following
requirements:
\begin{itemize}
\item the underlying topology or geometry is known;
\item the governing dynamics are physically or analytically well characterized;
\item ground truth is available either directly or through simulation;
\item the dataset is sufficiently large to test model robustness;
\item the computational cost is compatible with repeated experiments.
\end{itemize}

No single dataset is expected to be universal across all classes of
systems introduced in Section~\ref{sec-intro}.  Nevertheless, some
benchmark domains are particularly well suited for verifying models
based on gradient, curl, and topology-constrained flow decomposition.

We consider in primis water distribution networks, which are probably
the first source to look at, electrical power grids, good for
controlled generation of source/sink and loop-flow scenarios, and
fluid systems for which there is direct connection with the
Helmholtz-Hodge decomposition.
These examples are illustrative rather than exhaustive.

\subsection{Water Distribution Networks}

Water distribution networks represent the most attractive first
verification domain. They are network-constrained systems with known
topology, physically interpretable flows, sources, sinks, pressure
gradients, loops, and topology- dependent transport paths. Moreover,
hydraulic simulators such as EPANET \citep{EPANET} can generate
controlled scenarios in which the ground truth is known.

Two datasets are especially relevant. LeakDB \citep{LeakDB} is a
leakage diagnosis benchmark containing many artificially generated but
realistic leakage scenarios on water distribution networks, together
with scoring tools for algorithmic comparison.  DiTEC-WDN
\citep{DiTECWDN} is a larger spatiotemporal graph dataset containing
multiple water distribution networks and simulated hydraulic
scenarios.

These datasets are well suited for testing whether a model can recover
source-like anomalies, distinguish propagation from circulation, and
identify the topological features required to explain the observed
flow.

\subsection{Electrical Power Networks}

Electrical grids provide a second strong benchmark class. Their
topology is known, sources and sinks correspond to generators and
loads, and the governing equations are well established. Standard
benchmark systems are available through MATPOWER \citep{MATPOWER} and
PGLib-OPF \citep{PGLibOPF}.

Power networks are particularly useful for testing source-sink
balance, propagation over a fixed network, and loop-flow effects in
meshed grids. Compared with water networks, however, they may require
additional scenario generation to produce time-dependent datasets
suitable for training and behavioral verification.

\subsection{Continuous Fluid Systems}

Continuous fluid systems provide the most direct connection to the
classical Helmholtz-Hodge decomposition. Public simulation databases
such as the Johns Hopkins Turbulence Database \citep{JHTDB,DNS}
provide access to high-fidelity velocity and pressure fields generated
by direct numerical simulation.

These datasets are valuable for testing the continuous-to-discrete
transition, including the effects of discretization, mesh resolution,
and field sampling.  However, they are computationally heavier than
water or power-network datasets and are therefore better suited as a
second-stage benchmark.

\subsection{Lower-Priority Domains}

Traffic, pedestrian, and multi-agent trajectory datasets are useful
for testing models in noisy, data-driven, or controlled-agent
settings. Examples include vehicle trajectory datasets such as NGSIM
\citep{NGSIM} and multi-agent visual datasets such as the Stanford
Drone Dataset \citep{SDD}. These datasets are valuable for assessing
robustness and practical applicability, but they are less ideal for
primary verification because the ground truth mechanism generating the
observed dynamics is only partially known.

\subsection{Applicability to the Four System Classes}

Water distribution networks and electrical grids are therefore
recommended as primary verification domains. They offer the best
compromise between physical interpretability, accessibility, ground
truth, and computational tractability.  Continuous fluid datasets
should be used to test the connection with the underlying continuous
theory, while data-driven and multi-agent datasets should be reserved
for later validation and robustness studies.

This tiered strategy separates implementation verification from
application validation. A model should first demonstrate correct
behavior on controlled systems with known dynamics. Only then should
it be applied to target domains in which data are scarce, noisy,
sensitive, or only partially interpretable.

\begin{table}[h]
\caption{Candidate verification domains for the four classes of structured-flow systems.}
\label{tab:verification_domains}
  \centering
  \small
\begin{tabular}{llll}
\toprule
System class         & Best verification domain           & Ground truth & Main limitation \\
\midrule
Network-constrained  & Water networks, power grids        & Strong       & Domain-specific laws \\
Continuous systems   & Fluid simulations                  & Strong       & Computational cost \\
Data-driven networks & Epidemiology, finance, social  & Weak-medium & Ambiguous semantics \\
Multi-agent systems  & Traffic, swarms, orbits      & Medium       & Behavioral complexity \\
\bottomrule
\end{tabular}
\end{table}

\section{Conclusion}

This work has addressed the problem of modeling dynamical systems
governed by structured flows, characterized by gradient-driven,
curl-driven, and topology-constrained components. Starting from the
continuous formulation based on the Helmholtz-Hodge decomposition, we
have examined how such systems can be represented computationally
through discrete structures, with particular focus on the Graph Vector
Field (GVF) framework proposed by \citet{GVF2026}.

GVF provides a mathematically consistent and highly expressive
representation that unifies multimodal data and explicitly separates
the underlying mechanisms driving system dynamics. This decomposition
offers a strong interpretability advantage, enabling the
identification of source-driven effects, feedback loops, and global
topological constraints. However, this expressive power comes at the
cost of increased computational complexity, non-trivial topology
construction, and significant data requirements.

To address these limitations, we have proposed a hierarchy of
alternative modeling approaches, including parametric conditional
models, linear graph dynamical systems, and reduced Hodge
representations. These approaches trade expressivity for tractability,
offering practical solutions in settings where data are limited or
rapid inference is required. Importantly, we have shown that these
models can be interpreted as restricted instances of the GVF
framework, enabling a unified perspective across different levels of
complexity.

A central contribution of this work is the proposed strategy for model
assessment, based on a clear separation between structural
verification, behavioral verification, and domain-specific
validation. By leveraging datasets from well-understood physical
systems, it is possible to evaluate the correctness and robustness of
a model independently of the target application domain. This
cross-domain validation approach mitigates the risks associated with
data scarcity, noise, and ambiguity in real-world datasets.

The resulting framework supports an iterative modeling strategy in
which GVF may be used as a diagnostic tool to identify dominant
mechanisms, guiding the construction of reduced or parametric models
tailored to the problem at hand.  This approach enables a principled
trade-off between model fidelity, interpretability, and computational
efficiency.

Future work should focus on the systematic application of this
methodology to benchmark datasets, the quantitative comparison of
model classes under controlled conditions, and the extension of the
framework to domains with highly dynamic or partially observed
topologies. In particular, understanding the role of topology
construction and its impact on model performance remains a key open
question.

Overall, structured flow modeling provides a unifying perspective
across a wide range of domains, and the combination of expressive
frameworks such as GVF with pragmatic reduced-order models offers a
promising path toward scalable and interpretable data-driven modeling.

\appendix
\section{Unified Notation}\label{sec-unified}

Let $K$ be a simplicial complex with $n$ nodes, $e$ oriented edges,
and $f$ oriented faces. We denote by
$\mat{B}_1 \in \mathbb{R}^{n \times e}$ and
$\mat{B}_2 \in \mathbb{R}^{e \times f}$ the node-edge and edge-face
incidence matrices. With the convention
\begin{equation}
  \label{eq-ne-matrix}
  (\mat{B}_1)_{ve} =
  \begin{cases}
    -1, & \text{if edge } e \text{ leaves node } v \\
    +1, & \text{if edge } e \text{ enters node } v \\
    \;\;\;\; 0,  & \text{otherwise}
  \end{cases}
\end{equation}
the matrix produc $\mat{B}_1\,\vec{F}$ represents the net inflow
induced by the edge flow $\vec{F}$ at each node.

An edge-based flow field
\(
\vec{F}(t) \in \mathbb{R}^{e}
\)
can be decomposed as
\begin{equation}
  \label{eq-discrete-HHD-unified}
  \vec{F}(t) = \mat{B}_1^{\top}\,\vec{\varphi}(t) + \mat{B}_2\,\vec{\psi}(t) + \mat{H}\,\vec{\alpha}(t)
\end{equation}
where $\vec{\varphi}(t) \in \mathbb{R}^{n}$ is a scalar potential on nodes, 
$\vec{\psi}(t) \in \mathbb{R}^{f}$ is a 2-form on faces,
$\vec{\alpha}(t) \in \mathbb{R}^{b_1}$ with
the integer $b_1$ being the first Betti number of the complex,
and $\mat{H} \in \mathbb{R}^{e \times b_1}$ is a matrix whose columns span the
harmonic subspace. 

The harmonic subspace is defined as the kernel of the discrete 1-Hodge
Laplacian
\(
L_1 = \mat{B}_1^{\top}\mat{B}_1 + \mat{B}_2\mat{B}_2^{\top}
\)
such that
\(
\mathrm{im}(\mat{H}) = \ker L_1 
\).

A reduced Hodge model is obtained by replacing the full bases with
truncated ones as illustrated in section \ref{sec-HHD-reduced} and
resulting in equation \eqref{eq-HHD-reduced}, that is
\begin{equation}
  \label{eq-discrete-reduced-unified}
  \vec{F}(t) \approx \mat{U}_{\text{g,red}}\, \vec{a}_{\text{g}}(t)
                   + \mat{U}_{\text{c,red}}\, \vec{a}_{\text{c}}(t)
                   + \mat{U}_{\text{h,red}}\, \vec{a}_{\text{h}}(t),
\end{equation}
where the matrices 
$\mat{U}_{\text{g,red}} \in \mathbb{R}^{e \times K_{\text{g}}}$,
$\mat{U}_{\text{c,red}} \in \mathbb{R}^{e \times K_{\text{c}}}$,
and
$\mat{U}_{\text{h,red}} \in \mathbb{R}^{e \times K_{\text{h}}}$,
have reduced dimensionality compared to the full model: 
$K_{\text{g}} \ll n$,
$K_{\text{c}} \ll f$,
and
$K_{\text{h}} \leq b_1$.

Absorbing these three matrices into a single matrix
$\mat{W} \in \mathbb{R}^{e \times m}$ --- losing the decomposition
into gradient, curl, and harmonic components, and keeping only
pairwise interactions among a sufficiently high number of nodes ---
one gets the field representation used in a linear graph model:
\begin{equation}
  \label{eq-discrete-linear-unified}
  \vec{F}(t) = \mat{W}\,\vec{x}(t)
\end{equation}
where $\vec{x}(t)\in\mathbb{R}^{m}$ is an array representing the state
of $m$ 
vertices.\footnote{The number of vertices $m$ may be chosen in the
  range from $K_{\text{g}}$ to $n$ --- the lower the better --- to
  optimize the performance of the resulting model.}

The absence of simplicial complex structure means that this model does
not contain higher-order interactions that only exist when multiple
nodes are jointly active and cannot be reduced to a linear combination
of pairwise interactions.

High-order interactions are essential only when effects require joint
presence of multiple agents with context-dependent interactions and/or
group-level constraints.  Examples in which they are needed include
social contagion (triadic closure), biochemical reactions
(multi-molecule chemistry), and coordination in multi-agent systems.

Another possible way to reduce the dimensionality of the model without
adopting a linear graph representation is to encode the dynamics in a
parametric model.  When the underlying system is known, this allows to
encode the three components explicitly in a parametric form using
suitable functions:
\begin{equation}
  \label{eq-conditional-unified}
  \vec{F}(t;\vec{c}) = \mat{B}_{1}^{\top} \, g(\vec{x}(t);\vec{\theta},\vec{c})
                     + \mat{B}_{2} \, c(\vec{x}(t);\vec{\theta},\vec{c})
                     + \mat{H} \, h(\vec{x}(t);\vec{\theta},\vec{c})
\end{equation}
where \vec{\theta} represents an array of parameters learned or fitted
with the data, and \vec{c} includes other context parameters that are
observed or prescribed and act as external inputs to the model.

\section{Dynamic Evolution for Continuous Systems}\label{app-dynamic}

To obtain a time-dependent dynamical system, the edge flow must be
coupled to a node-state evolution equation. In the continuous case,
the continuity equation for a locally conserved quantity $q$ is
\begin{equation}
  \frac{\partial \vec{x}}{\partial t} + \nabla \cdot \vec{J} = \vec{s}
\end{equation}
where $\vec{J} = q \vec{v}$ is the flux of quantity $q$ with velocity
\vec{v} and $s$ is an external generator of $q$ per unit time and per
unit volume (i.e.\ a source or a sink of $q$).  This is the same as
equation \eqref{eq-continuity}.
With the node-edge incidence matrix defined by equation
\eqref{eq-ne-matrix} above, the corresponding discrete-time evolution
is
\begin{equation}
  \frac{\de\vec{x}}{\de t} = \mat{B}_1 \, \vec{F}(t) + \vec{s}(t),
\end{equation}
or, after time discretization,
\begin{equation}
  \label{eq-time-discrete-continuity}
  \vec{x}(t+\Delta t) = \vec{x}(t) + \Delta t\,\mat{B}_1\, \vec{F}(t) + \Delta t\,\vec{s}(t)
\end{equation}
which is the same as equation \eqref{eq-time-evol} above.  Here,
\vec{F} may be represented in the form \eqref{eq-discrete-HHD-unified}
for GVF, as \eqref{eq-discrete-reduced-unified} for a reduced Hodge
model, as \eqref{eq-discrete-linear-unified} for a linear graph model,
or as \eqref{eq-conditional-unified} in the case of a parametric
conditional model.

If the underlying domain is inherently continuous, the operators
approximate differential operators, and the functions converge to
smooth fields, then as the mesh size goes to zero in a parametric
conditional model equation \eqref{eq-time-evol} essentially becomes a
partial differential equation encapsulating both time evolution and
conservation.

Note that the GVF formulation provides a decomposition of a discrete
flow field, but does not by itself define a partial differential
equation. The latter arises only when the flow is coupled with an
evolution law, typically through a conservation equation.

Inserting equation \eqref{eq-discrete-HHD-unified} into
\eqref{eq-time-evol} and remembering that the divergence of a curl is
null (the discrete version is $\mat{B}_1\,\mat{B}_2=0$) and that the
harmonic component is divergence-free (or $\mat{B}_1\,\mat{H}=0$), we
get the discrete analog of a partial differential equation:
\begin{equation}
  \begin{split}
    \vec{x}(t+\Delta t) &= \vec{x}(t) + \Delta t\,\{\mat{B}_1\, [\mat{B}_1^{\top}\,\vec{\varphi}(t) + \mat{B}_2\,\vec{\psi}(t) + \mat{H}\,\vec{\alpha}(t)] + \vec{s}(t)\}
    \\
                        &= \vec{x}(t) + \Delta t\, [\mat{B}_1\, \mat{B}_1^{\top}\,\vec{\varphi}(t) + \vec{s}(t)]
    \\
                        &= \vec{x}(t) + \Delta t\, [\mat{B}_1\, \vec{F}_{\mathrm{grad}} + \vec{s}(t)]
  \end{split}
\end{equation}
where $\mat{B}_1\, \mat{B}_1^{\top}$ is the discrete Laplacian
operator and the last result is obtained by inserting equation
\eqref{eq-DEC-grad}.  Therefore, it is the gradient component
$\vec{F}_{\mathrm{grad}}$ of our vector field that drives the dynamics
of a system obeying a continuity equation, as expressed in equation
\eqref{eq-time-discrete-continuity}.

\section{Introducing Metric Information}\label{sec-metric}

In weighted or geometric settings, the Euclidean inner products are
replaced by mass matrices or discrete Hodge star operators. The
incidence matrices still encode the topology, while the Hodge star
operators encode the metric properties of the system.

Let $\mat{M}_0 \in \mathbb{R}^{n \times n}$,
$\mat{M}_1 \in \mathbb{R}^{e \times e}$, and
$\mat{M}_2 \in \mathbb{R}^{f \times f}$ be positive definite mass
matrices (discrete Hodge star operators) defining the inner products
on node-, edge-, and face-based quantities, respectively. The
incidence matrices $\mat{B}_1$ and $\mat{B}_2$ remain purely
topological, with entries in $\{-1,0,+1\}$, while geometry and
physical properties are encoded in $\mat{M}_0$, $\mat{M}_1$, and
$\mat{M}_2$ as metric information such as lengths, areas, volumes,
conductances, or capacities.  They are typically diagonal.

The weighted Helmholtz-Hodge decomposition of an edge field becomes:
\begin{equation}
  \vec{F} = \mat{B}_1^{\top}\,\vec{\varphi} + \mat{M}_1^{-1} \,\mat{B}_2 \, \mat{M}_2 \, \vec{\psi} + \vec{h}
\end{equation}
with the harmonic component defined by the weighted conditions
\begin{equation}
  \mat{B}_1 \,\mat{M}_1 \, \vec{h} = 0,
  \qquad
  \mat{B}_2^{\top} \, \vec{h} = 0.
\end{equation}
Equivalently, \vec{h} belongs to the kernel of the weighted 1-Hodge Laplacian:
\begin{equation}
  L_1^{(w)} = \mat{M}_1^{-1} \, \mat{B}_1^{\top} \, \mat{M}_0 \, \mat{B}_1 + \mat{M}_1^{-1} \, \mat{B}_2 \, \mat{M}_2 \, \mat{B}_2^{\top}.
\end{equation}

The node weights in $\mat{M}_0$ represent volume around node (fluid
system), or storage capacity (heat capacity in thermal systems;
population size in epidemiology).  Edge weights in $\mat{M}_1$
represent how easily flow moves along edges, like electrical
conductance, hydraulic conductance, or mobility (diffusion).  Face
weights in $\mat{M}_2$ represent area (face area in fluid; surface
element in electromagnetic system) or strength surfaces (coupling
between loops in transport systems).

In practice, the metric ($\mat{M}_0$, $\mat{M}_1$, and $\mat{M}_2$)
tells the decomposition which flows are ``expensive'' or ``easy''
along different edges, while topology ($\mat{B}_1$ and $\mat{B}_2$)
defines what is connect to what.


\begin{thebibliography}{14}
\providecommand{\doi}[1]{\href{https://doi.org/#1}{doi: #1}}
\providecommand{\arxiv}[1]{\href{https://arxiv.org/abs/#1}{arXiv: #1}}

\bibitem[Coletti and Fallucchi(2026)]{GVF2026}
Silvano Coletti and Francesca Fallucchi.
\newblock {Graph Vector Field}: A unified framework for multimodal health risk
  assessment from heterogeneous wearable and environmental data streams, 2026.
\newblock \arxiv{2603.28115}.

\bibitem[Hatcher(2002)]{Hatcher2002}
Allen Hatcher.
\newblock \emph{Algebraic Topology}.
\newblock Cambridge University Press, 2002.

\bibitem[Hughes(1987)]{Hughes1987}
Thomas J.~R. Hughes.
\newblock \emph{The Finite Element Method: Linear Static and Dynamic Finite
  Element Analysis}.
\newblock Prentice Hall, 1987.

\bibitem[Bathe(2014)]{Bathe1995}
Klaus-Jürgen Bathe.
\newblock \emph{Finite Element Procedures, 2nd Ed.}
\newblock Pearson Education, 2014.

\bibitem[Lim(2020)]{Lim2020}
Lek-Heng Lim.
\newblock {Hodge Laplacians on Graphs}.
\newblock \emph{SIAM Review}, 62\penalty0 (3):\penalty0 685--715, 2020.
\newblock \doi{10.1137/18M1223101}.

\bibitem[{US Environmental Protection Agency}()]{EPANET}
{US Environmental Protection Agency}.
\newblock {EPANET: Application for Modeling Drinking Water Distribution
  Systems}.
\newblock URL \url{https://www.epa.gov/water-research/epanet}.

\bibitem[Vrachimis et~al.(2018)Vrachimis, Kyriakou, Eliades, and
  Polycarpou]{LeakDB}
S.~G. Vrachimis, M.~S. Kyriakou, D.~G. Eliades, and M.~M. Polycarpou.
\newblock {LeakDB: A benchmark dataset for leakage diagnosis in water
  distribution networks}.
\newblock In \emph{{Proc. of WDSA / CCWI Joint Conference}}, volume~1, 2018.
\newblock URL \url{https://github.com/KIOS-Research/LeakDB}.

\bibitem[Truong et~al.(2025)Truong, Tello, Lazovik, and Degeler]{DiTECWDN}
Huy Truong, Andrés Tello, Alexander Lazovik, and Victoria Degeler.
\newblock {DiTEC-WDN: A Large-Scale Dataset of Hydraulic Scenarios across
  Multiple Water Distribution Networks}, 2025.
\newblock URL \url{https://huggingface.co/datasets/rugds/ditec-wdn}.

\bibitem[MAT()]{MATPOWER}
{MATPOWER: Power System Simulation Package}.
\newblock URL \url{https://matpower.org/}.

\bibitem[PGL()]{PGLibOPF}
{Power Grid Lib: Benchmarks for Optimal Power Flow}.
\newblock URL \url{https://github.com/power-grid-lib/pglib-opf}.

\bibitem[JHT()]{JHTDB}
{Johns Hopkins Turbulence Database}.
\newblock URL \url{https://turbulence.idies.jhu.edu/home}.

\bibitem[Li et~al.(2008)Li, Perlman, Wan, Yang, Meneveau, Burns, Chen, Szalay,
  and Eyink]{DNS}
Yi~Li, Eric Perlman, Minping Wan, Yunke Yang, Charles Meneveau, Randal Burns,
  Shiyi Chen, Alexander Szalay, and Gregory Eyink.
\newblock {A public turbulence database cluster and applications to study
  Lagrangian evolution of velocity increments in turbulence}.
\newblock \emph{Journal of Turbulence}, 9:\penalty0 1--29, 2008.
\newblock ISSN 1468-5248.
\newblock \doi{10.1080/14685240802376389}.

\bibitem[{U.S. Department of Transportation Federal Highway
  Administration}(2016)]{NGSIM}
{U.S. Department of Transportation Federal Highway Administration}.
\newblock {Next Generation Simulation (NGSIM) Vehicle Trajectories and
  Supporting Data}, 2016.
\newblock \doi{10.21949/1504477}.

\bibitem[Robicquet et~al.(2016)Robicquet, Sadeghian, Alahi, and Savarese]{SDD}
A.~Robicquet, A.~Sadeghian, A.~Alahi, and S.~Savarese.
\newblock Learning social etiquette: Human trajectory prediction in crowded
  scenes.
\newblock In \emph{European Conference on Computer Vision (ECCV)}, 2016.
\newblock URL \url{https://cvgl.stanford.edu/projects/uav\_data/}.
\newblock Stanford Drone Dataset (SDD).

\end{thebibliography}

\end{document}